\documentclass[]{aa} %\includegraphics*[width=10.1cm]{fig2_fit.ps}
\usepackage{psfig}                          
\newcommand{\be}{\begin{equation}} 
\newcommand{\en}{\end{equation}}

\def\zabs{$z_{\rm abs}$}

\def\h2{H$_2$}

\def\feii{Fe~{\sc ii}~}
\def\feiia{Fe~{\sc ii}$\lambda$1608~}  
 
\def\feiistr{Fe~{\sc ii}$\lambda\lambda$2383,2600~} 

\def\mgii{Mg~{\sc ii}~}

\def\kms{km~s$^{-1}$~} 
 \def\chisq{$\chi^{2}$~} 
\def\dela{$\Delta\alpha/\alpha$~} 
\usepackage{txfonts}
\begin{document}
\title{On the variation of the fine-structure constant: 
Very high resolution spectrum of QSO HE {0515$-$4414}
%High resolution analysis of \zabs=1.1508
% toward HE {0515-4414} using HAPPS/ESO}
%%    \author{G. Wuchterl
%%           \inst{1}
%%           \and
%%           C. Ptolemy\inst{2}\fnmsep\thanks{Just to show the usage
%%           of the elements in the author field}
%%           }
\author{Hum Chand\inst{1}, 
    Raghunathan Srianand\inst{1},
    Patrick Petitjean\inst{2,3},\\
    Bastien Aracil\inst{2,4},
    Ralf Quast\inst{5},
    Dieter Reimers\inst{5}
} 
\thanks{ 
Based on observations collected 
at the European Southern Observatory (ESO), under Programe   
ID No. 072.A-0244 with HARPS on the 3.6~m 
telescope operated at the La Silla Observatory and  Programe
ID 066.A-0212 with
UVES/VLT at the Paranal observatory.}
}  
\titlerunning{Variation of the fine-structure constant}
\authorrunning{H. Chand et al.}  
\author{Hum Chand\inst{1}, 
    Raghunathan Srianand\inst{1},
    Patrick Petitjean\inst{2,3},\\
    Bastien Aracil\inst{2,4},
    Ralf Quast\inst{5},
    Dieter Reimers\inst{5}
} 
\offprints{H. Chand \\~\email{hcverma@iucaa.ernet.in}} 
\institute{$^1$IUCAA, Post Bag 4,
  Ganeshkhind, Pune 411 007, India\\ 
$^2$Institut d'Astrophysique de Paris,
UMR7095 CNRS, Universite Pierre \& Marie Curie,
98 bis boulevard Arago,
75014 Paris.\\
$^3$LERMA, Observatoire de Paris,
  61 Rue de l'Observatoire, 
  F-75014 Paris, France\\ 
$^4$Department of Astronomy,
   University of Massachusetts, 
   710 North Pleasant Street,
   Amherst, MA 01003-9305, USA\\ 
$^5$Hamburger Sternwarte, Universitat Hamburg, Gojenbergsweg 112,
D-21209 Hamburg, Germany\\
} 
\date{Received date/ Accepted date} 
%%%%%%%%%%%%%%%%%%%%%%%%%%%%%%%%%%%%%%%%%%%%%%%%%%%%%%%%%%%%%%%%%%%%%%%%
\abstract
{}
%aim
{We present a detailed analysis of a very high resolution
(R~$\approx 112,000$) spectrum of  the quasar HE {0515$-$4414} 
obtained using the High Accuracy Radial velocity Planet Searcher 
(HARPS) mounted on the ESO 3.6~m telescope at the La Silla observatory. 
The main aim is to use HARPS spectrum of very high 
wavelength calibration accuracy (better than 1~m\AA), to constrain
the variation of $\alpha=e^2/\hbar c$ and investigate 
any possible systematic inaccuracies in the 
wavelength calibration of the UV Echelle Spectrograph (UVES) mounted on the ESO Very
Large Telescope (VLT).} 
%method
{A cross-correlation analysis between the Th-Ar lamp 
spectra obtained with HARPS and UVES is carried  out
to detect any possible shift between the two spectra.
Absolute wavelength calibration accuracies, and how that translate 
to the uncertainties in \dela are computed using Gaussian fits 
for both lamp spectra. The value of \dela  at \zabs~=~1.1508 is obtained 
using Many Multiplet method, and simultaneous
Voigt profile fits of HARPS and UVES spectra.}
%results
{We find the shift between the HARPS and UVES spectra has mean around zero with a dispersion 
of $\sigma\simeq 1$ m\AA.
This is shown to be well within the wavelength calibration accuracy of UVES
(i.e $\sigma\simeq 4$ m\AA). We show that the uncertainties in the 
wavelength calibration induce an error of about, 
\dela $~\le 10^{-6}$, in the determination of the
variation of the fine-structure constant.
Thus, the results of non-evolving \dela reported in the 
literature based on UVES/VLT data should not be heavily influenced  by 
problems related to wavelength calibration uncertainties.
Our higher resolution spectrum 
of the \zabs~=~1.1508 Damped Lyman-$\alpha$ system toward HE {0515$-$4414} 
reveals more components compared to the UVES spectrum.
Using only \feii lines of  \zabs~=~1.1508 system, we obtain
\dela~=~${ (0.05\pm0.24)\times10^{-5}}$. This result is consistent with
the earlier measurement for this system using the UVES spectrum alone.}
{}
\keywords{ 
{\em Quasars:} absorption lines -- 
{\em cosmology:} observations 
} 
\maketitle 
%%%%%%%%%%%%%%%%%%%%%%%%%%%%%SECTION 1%%%%%%%%%%%%%%%%%%%%%%%%%%% 
\section{Introduction}
%%%%%%%%%%%%%%%%%%%%%%%%%%%%%%%%%%%%%%%%%%%%%%%%%%%%%%%%%%%%%%%%%
\label{sect:Int}
%%%%
Some of the modern theories of fundamental physics, such as SUSY,
GUT and Super-string theory, allow possible space and time variations of 
the fundamental constants, thus motivating an
experimental search for such a variation (Uzan 2003 and 2004 
for a detail review on the subject).  Murphy et al. (2003), 
applying the Many Multiplet method (MM method) to 143
complex metal line systems, claimed
a non-zero variation of the fine-structure constant, $\alpha=e^2/\hbar c$:
$\langle\Delta\alpha/\alpha\rangle = (-0.57\pm0.11)\times10^{-5}$ 
for $0.2\le z \le 3.5$,
where $\Delta\alpha/\alpha=(\alpha_{z}-\alpha_{0})/\alpha_{0}$, with
$\alpha_{0}$ being the present value and $\alpha_{z}$ its value at redshift $z$.
This result, if true, would have very 
important implications to our understanding of fundamental
physics and has therefore motivated new activities in the field.
Search for the possible time-variation of $\alpha$ using alkali doublets
has started long ago (Bahcall et al. 1967).
The alkali-doublet method is a clean method for constraining
the variation in $\alpha$ using spectral lines because it uses transitions
from the same species
(Wolfe et al. 1976; Levshakov 1994; Potekhin et al. 1994; Cowie \&
Songaila, 1995; Varshalovich et al. 1996; Varshalovich et al. 2000; 
Murphy et al. 2001a; Martinez et al. 2003; Chand et al. 2005). 
The tightest constraint obtained using this method till date 
is \dela = (0.15$\pm$0.44) $\times10^{-5}$ at 
$z \sim 2$ (Chand et al. 2005).
\par
Studies based on heavy element molecular absorption lines seen in the
radio/mm wavelength range are more sensitive than that
based on optical/UV absorption lines. 
They usually provide constraints on the variation of a combination of 
the fine-structure constant, the proton g-factor ($G_{p}$) and 
the electron-to-proton mass ratio ($\mu$). Murphy et al. (2001b)
have obtained \dela = $(-0.10\pm0.22)\times10^{-5}$ at 
z = 0.2467 and \dela = $(-0.08\pm0.27)\times
10^{-5}$ at z = 0.6847, assuming a constant proton g-factor ($G_{p}$). 
It has been pointed out that OH lines are very useful in
simultaneously constraining various fundamental constants
(Chengalur \& Kanekar 2003; Kanekar et al. 2004;
Darling 2003, 2004). These studies have provided 
\dela = $(0.6\pm1.0)\times 10^{-5}$ for an absorption system
at \zabs = 0.247 toward PKS~1413+135. Such
studies have not been performed yet at higher redshift (i.e $z$~$\ge1$) 
due to the lack of molecular absorption systems. \par
%%%%%%%%%
Constraints on the variations of $\alpha$ are also obtained from
terrestrial measurements.
The most stringent constrain has been obtained from the analysis
of the Oklo phenomenon. Fujii et al. (2000)  find that $\Delta\alpha/\alpha = 
(-0.8\pm1.0)\times10^{-8}$
over a period of about 2 billion years (or $z\simeq0.45$).
Laboratory experiments also give very stringent constraints on the
local variation of $\alpha$. Marion et al. (2003) have obtained, $\Delta\alpha/\alpha\Delta t =
(-0.4\pm16)\times10^{-16}\,{\rm yr}^{-1}$, by
comparing the hyperfine transition in $^{87}$Rb and  $^{133}$Cs over a
period of 4 years assuming no variation in the magnetic moments.  
Fischer et al. (2004) have obtained, $\Delta\alpha/\alpha\Delta t =
(-0.9\pm2.9)\times10^{-16}\,{\rm yr}^{-1}$, by comparing the absolute
$1S-2S$ transition of atomic hydrogen to the ground state of
Cesium. A linear extrapolation gives a constraint of 
$-1.3\times10^{-6}\le$ \dela$\le1.9\times10^{-6}$ at $z = 1$ for 
the most favored cosmology ($\Omega_m = 0.27$, $\Omega_{\Lambda} =0.73$ and $ h=0.71$).
\par
Clearly all the experimental results summarized above are
consistent with no variation of $\alpha$. However, these results
do not directly conflict with the positive detection by Murphy et al. (2003)
either because of the insufficient sensitivity of the method (as in the case of 
alkali doublets) or because of  the different redshift coverage (as in 
the case of radio and terrestrial measurements). 
However,  recent attempts using the MM method (or its modified version)
applied to very high quality UVES spectra have resulted in null detections.
The analysis of Fe~{\sc ii} multiplets and Mg~{\sc ii} doublets
in a homogeneous sample of 23 systems has yielded a stringent constraint, 
\dela = $(-0.06\pm0.06)\times10^{-5}$ (Chand et al. 2004; 
Srianand et al. 2004). Modified MM method analysis of \zabs = 1.1508 
toward HE 0515$-$4414 that avoids possible complications due to isotopic 
abundances has resulted 
in \dela = $(0.01 \pm 0.17)\times 10^{-5}$ (Quast et al. 2004). 
Levshakov et al. (2005b) have  
re-analysis this system using the single ion differential
alpha measurement method as described in Levshakov et al. (2005a), 
and obtained \dela = $(-0.007\pm0.084)\times10^{-5}$.
Clearly all studies based on VLT-UVES data are in contradiction with the conclusions
of Murphy et al. (2003). 
\par
A first possible concern about these studies is the real
accuracy and robustness of the various calibration procedures. 
A second possible source of uncertainty comes from the multi-component Voigt-profile
decomposition. 
It is very important to check how sensitive the derived constraints are to
the profile decomposition. This can be done by performing the analysis on data 
of higher resolution than typical UVES (or HIRES) spectra.
The best way to investigate all this is to compare data taken by 
UVES (or HIRES) with data on the same object taken with another
completely independent, well controlled, and higher spectral resolution instrument.
The advent of HARPS mounted on the ESO 3.6~m telescope makes this possible.
Unfortunately this is only possible on the brightest quasar in the
southern sky, HE~0515$-$4414.
\par
This forms the basic motivations of this work.
We report the analysis of the \zabs = 1.15 DLA system
toward QSO HE 0515$-$4414 (De la Varga et al. 2000, Quast et al. 2004, 2005)
using very high resolution (R$\sim112,000$) spectra obtained with
HARPS mounted on the ESO 3.6~m telescope.
The organization of the  paper is  as follows. The HARPS
observations of HE 0515$-$4414 are described in  Section 2.
Calibration accuracy and comparison with the UVES observations
are discussed in Section 3. In Section 4 we present the 
joint analysis of the HARPS and UVES spectra. Results 
are summarized and discussed  in Section 5.
%%%%%%%%%%%%%%%%%%%%% SECTION 2%%%%%%%%%%%%%%%%%%%%%%%%%
\section{Observations} 
%%%%%%%%%%%%%%%%%%%%%%%%%%%%%%%%%%%%%%%%%%%%%%%%%%%%%%%%%%
\label{sect:Obs}
%%%%
The spectrum of HE 0515$-$4414 used in this work was obtained with the 
High Accuracy Radial velocity Planet Searcher (HARPS)
mounted on the ESO 3.6~m telescope at the La Silla observatory.     
HARPS is a fiber-fed spectrograph and is therefore less affected by
any fluctuation in the seeing conditions (Mosser et al. 2004).
It is installed in the Coud\'e room of the 3.6~m telescope building and is
enclosed in a box in which vacuum and constant temperature are maintained.
The instrument has been specifically designed to guarantee stability
and high-accuracy wavelength calibration.
\par
The observations were carried over four nights in classical
 fiber spectroscopy mode, with one fiber on the target and
 the other on the sky.
 The CCD was read in normal low readout mode without binning.
 The echelle order extraction from the raw data frame is done
 using the HARPS reduction pipeline. 
 The error spectrum is computed by modeling the photon
 noise with a Poisson distribution and 
 CCD readout noise with a Gaussian distribution.  
  The calibrated spectrum is converted to vacuum wavelengths according
 to Edl\'en (1966) and the heliocentric velocity correction is done 
 manually using the dedicated MIDAS (ESO-Munich Image Data Analysis Software) procedure.
Special attention was given while merging the orders.
While combining overlapping regions, higher weights were assigned to the 
wavelength ranges toward the
center of the order compared to the one at the edges.
 The resulting 1-D spectrum covers the wavelength range from 3800 to 6900
 \AA, with a gap between 5300 to 5330 \AA~ caused by the transition between 
the two CCDs used in HARPS. 
In total, we obtained 14 individual exposures, each
of duration between 1 and 1.5 hour. Combination of individual exposures
is performed using  a sliding window and weighting the signal by the
errors in each pixel.  
The final error spectrum was obtained 
by adding quadratically in each pixel the extracted errors 
and the rms of the 14 individual measurements. The final
combined spectrum has a S/N ratio of about 30 to 40 per pixel
of size $\sim$0.015 \AA~ and  a spectral  resolution of  R$\approx$ 112,000.\par
 %%%
 To make quantitative comparisons, as will be discussed in
 the next section, we have also used the UVES spectrum of this QSO.
 The details of the UVES observation and data reduction can be found
 in Quast et al. (2004). However we have used our procedures  for air-to-vacuum
 wavelength conversion, heliocentric velocity correction 
and for the addition of individual exposures
as in the case of the HARPS spectrum.
%%%%%%%%%%%%%%%%%%%%%%%%%SECTION 3%%%%%%%%%%%%%%%%%%%%%%%%%%
\section{Accuracy of wavelength calibration}
%%%%%%%%%%%%%%%%%%%%%%%%%%%%%%%%%%%%%%%%%%%%%%%%%%%%%%%%%%%
\label{sect:comp_cal}
In this Section we investigate (i) the cross-correlation between
the Th-Ar lamp spectra obtained with HARPS and UVES, (ii) the absolute wavelength 
calibration accuracies of HARPS and UVES and (iii) how the uncertainties 
in the wavelength calibration translate into uncertainties in \dela measurements 
in the case of HARPS and UVES.
\subsection{Cross-correlation of UVES and HARPS Th-Ar spectra}
\label{crossCorr:subsec}
To estimate how well the UVES and HARPS wavelength scales agree, 
one can in principle use the narrow heavy element absorption lines
seen in the spectra of the QSO. 
However not only the number of such lines is small but also,
due to differences in the resolutions and S/N ratios, spurious
shifts can be introduced in the analysis. In order to
avoid this, we perform a cross-correlation analysis between the 
Th-Ar lamp spectra obtained with UVES and HARPS.
We have 4 and 14 Th-Ar lamp exposures respectively for UVES 
and HARPS observations in the setting that covers the 
wavelength range where Fe~{\sc ii} and Mg~{\sc ii} absorption lines from
the \zabs~=~1.1508 absorption system are seen.
We have combined
all the extracted Th-Ar exposures after subtracting a smooth
continuum corresponding to the background light. 
\par
\begin{figure}[h]
\psfig{figure=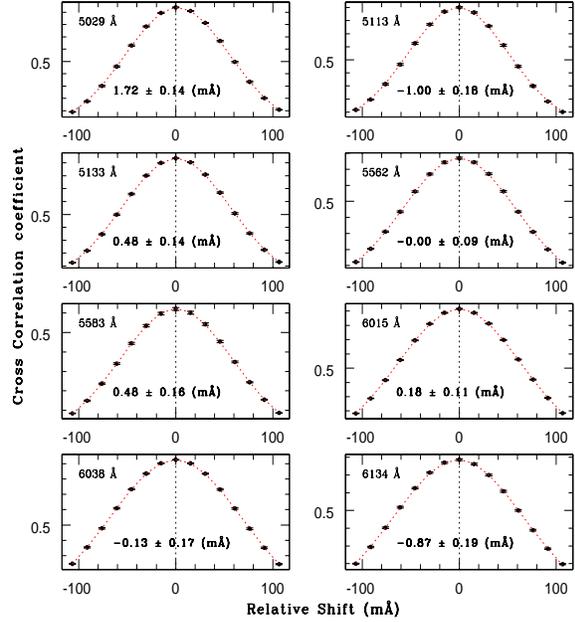,height=9.cm,width=8.1cm,angle=00} 
\caption[]{The points with error-bars are the cross-correlation coefficients plotted as a  
 function of the relative shift between UVES and HARPS Th-Ar spectra.
 The cross-correlation is performed using 
groups of five consecutive unblended emission lines 
 in the vicinity of the different metal absorption lines of  the \zabs~=~1.1508 system.
 The observed wavelength of the region around the metal line is given in the 
 top left corner in each panel.
 The dotted line is the best Gaussian fit to these coefficients. This is used to derive the 
 relative shift between the two spectra with sub-pixel accuracy.
 The mean relative shift and 1$\sigma$ error as well as
 the central wavelength of the region used are given in each panel.}
 \label{crossc.fig} 
\end{figure}

The cross-correlation analysis was performed 
on groups of five consecutive unblended emission lines that 
are clearly seen in both the UVES and HARPS spectra. 
For this, both spectra were re-sampled to an uniform wavelength scale using  
cubic spline and the pixel-by-pixel cross correlation was performed 
by shifting the UVES spectrum with respect to the HARPS spectrum. The results of the
cross-correlation at places where absorption lines at \zabs~=~1.1508 are 
redshifted are shown in Fig.~\ref{crossc.fig}. 
All the curves shown in this figure have their peak at zero pixel shift
with a typical pixel size of 15 m\AA.
In order to derive sub-pixel
accuracy in the cross-correlation, we have fitted a Gaussian to 
the cross correlation curves as is shown by dotted lines 
(Fig.~\ref{crossc.fig}) and derive its centroid accurately.
The corresponding values are given in each panel. The
relative shifts between the two spectra are less than 1 m\AA~
except in one case where it is 1.7~m\AA. We note that the quadratic refinement
technique (instead of a Gaussian fitting) also gives similar results.
%%%%
To derive the global trend of the relative shift, we have extended our 
cross-correlation analysis, to the entire wavelength range.
%%%
\begin{figure}
\psfig{figure=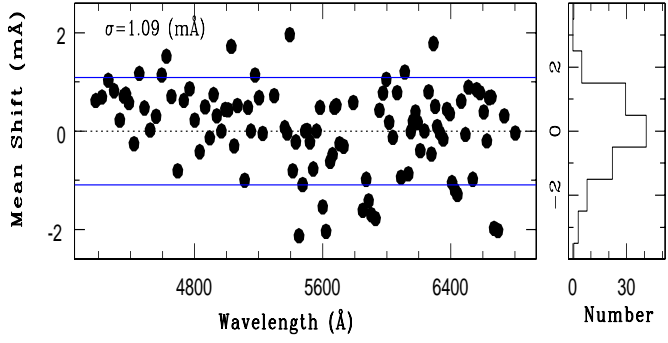,height=5.cm,width=9.cm,angle=00,bbllx=21bp,bblly=150bp,bburx=573bp,bbury=360bp,clip=true} 
\caption{
 The left panel shows the shift between HARPS and UVES lamp spectra 
 derived by performing  the cross-correlation (as shown in Fig.~\ref{crossc.fig}) over the  
 wavelength regions consisting of five consecutive unblended Th-Ar lines. The histogram of
 the mean shift is shown in the right panel. The mean shift is 0.01~m\AA~ and the rms
$\sigma$~=~1.09~m\AA.}
\label{crossHARPSUV.fig}
\end{figure} 
%%%
The result of the analysis is shown in Fig.~\ref{crossHARPSUV.fig}. 
The shifts are obtained in the same way as in Fig.~\ref{crossc.fig}.
The average of the mean relative shifts 
over the entire wavelength range is 0.01 m\AA~ with an rms deviation of 1.09 m\AA. 
In what follows we investigate the absolute wavelength calibration accuracies of 
the two instruments.
%%%%%%%%%%%%%%%
\subsection{Testing absolute wavelength calibration error of UVES and HARPS}
\label{abscal:subsec}
%%%%%%%%%%%%%%%%
To test the absolute wavelength calibration accuracy we compare
the central wavelength of strong un-blended emission lines in
the extracted Th-Ar lamp spectrum with the wavelengths tabulated in Cuyper
et al. (1998). We model the emission lines by a single Gaussian function.
The best-fit line-centroid along
with other parameters of the models and errors are determined by a
\chisq minimization procedure. In many cases we 
find it difficult to fit the lines with  reduced $\chi^{2}\approx 1$.
%%%%%
%
\begin{figure}
\psfig{figure=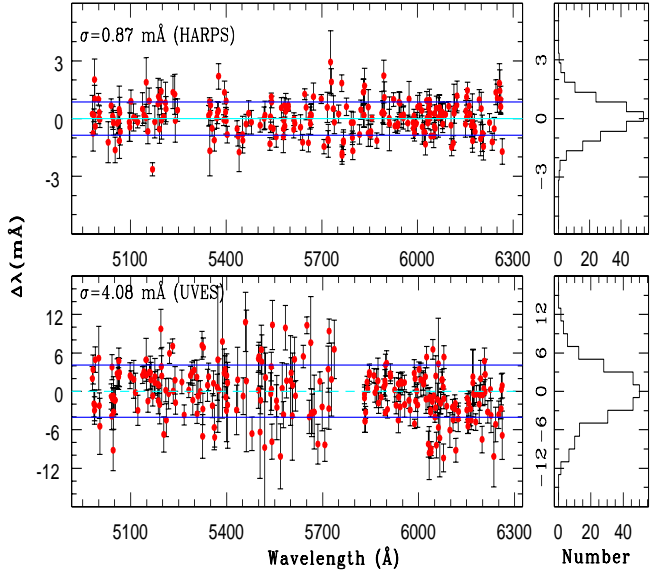,height=8.cm,width=9.cm,angle=00,bbllx=22bp,bblly=160bp,bburx=587bp,bbury=533bp,clip=true}
\caption[]{$\Delta\lambda$, the offset of the 
centroid wavelength of the emission lines in the Th-Ar lamp 
spectra (obtained using Gaussian fits) with respect to the 
wavelengths given by Cuyper et al. (1998) is plotted versus wavelength. 
The left-hand side upper and lower panels
show the results for the  HARPS and UVES spectra respectively. 
The corresponding right-hand side panels provide histograms.  
The  root mean square deviation of $\Delta\lambda$ around zero ($\sigma$) is
stated explicitly.
}
\label{dlam.fig} 
\end{figure} 
%%%%%
 In such cases we have scaled the flux errors by square root 
 of the reduced \chisq and re-run the fitting procedure. In this way, we 
 have avoided any underestimation of the errors on the best fit parameters,
 assuming that the actual errors on the flux of the Th-Ar lamp spectrum
 was somehow underestimated.\par  
 The difference between the best-fit line centroid,  
in the extracted lamp spectra and the wavelength quoted by Cuyper et al. (1998) 
is plotted in Fig.~\ref{dlam.fig}. The wavelength range shown in this
figure is the one covered by the main \feii and \mgii lines of the
\zabs~=~1.1508 system. We find the rms of the deviation
($\Delta\lambda$ in Fig.~\ref{dlam.fig})  around zero to be,
respectively, 0.87~m\AA~ and 4.08~m\AA~ for the HARPS and UVES lamp spectra.
This clearly demonstrates that the shifts between the HARPS and the 
UVES lamp spectra measured from the cross-correlation analysis  (i.e 
$\le1$ m\AA) are well within the wavelength calibration accuracy
of UVES.\par
\begin{figure}
\psfig{figure=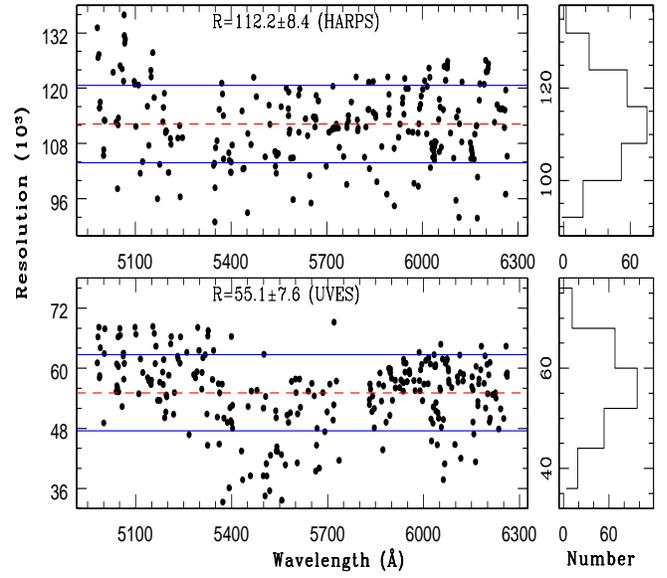,height=8.cm,width=9.cm,angle=00,bbllx=22bp,bblly=160bp,bburx=587bp,bbury=533bp,clip=true}
\caption[]{ 
 The spectral resolution ($R=\lambda/FWHM$) measured from Gaussian fitting  of emission 
lines in the Th-Ar lamp spectra is plotted versus wavelength.
The left-hand side upper and lower panels show the result from  HARPS and UVES spectra respectively. 
The corresponding histograms are shown in the right-hand side panels.  
The mean along with the standard deviation are stated explicitly.
}
\label{reso.fig} 
\end{figure}

 
In addition, we have used the best-fit FWHM of the Gaussian fit of the lamp lines
to derive the spectral resolution ($R=\lambda/FWHM$) of the spectrum. The 
resolution measurements are shown in Fig.~\ref{reso.fig}. The mean resolution
 and standard deviation for HARPS and UVES 
are found to be $R= 112,200$ and $\sigma=8,400$; $R= 55,100$ and $\sigma=7,600$ respectively.\par
%%
%%%%%%%
%%%
%%%%%%
%
\subsection{Effect of calibration error on $\Delta\alpha/\alpha$
 measurement}
\label{calDal:subsec}
Next we investigate  how the scatter in wavelength calibration
($\Delta\lambda$) translates into a scatter in \dela. 
We follow the method used by Murphy et al. (2003) for this purpose.
We randomly choose 3 emission lines in the
lamp spectrum, with a rest wavelength close to each of the observed wavelengths
of the \feii and \mgii lines used in the analysis of the variation of $\alpha$.
There are two \mgii lines, $\lambda$2796 and $\lambda$2803, and
five \feii lines, $\lambda$2344, $\lambda$2374, $\lambda$2382, $\lambda$2586, 
and $\lambda$2600. Thus we have 21 ($7\times3$) lines per realization.
By choosing 3 lines,  we mimic 3 distinct components in the actual absorption system. 
We assume that the measured shift in the emission line centroid away from
the actual value is caused by the variation in $\alpha$.
To estimate this variation, we use the analytic fitting function given by 
Dzuba et al. (2002),
\begin{equation}
w = w_o + q x.
\end{equation} 
Here, $w_o$ and $w$ are, respectively, the vacuum wave number
(in units of cm$^{-1}$) measured in the laboratory and 
the modified wave number due to a change in $\alpha$;
$x=(\Delta\alpha/\alpha+1)^2-1$ and $q$ is the
sensitivity coefficient.
At each chosen lamp emission line we assign the $q$ value of the neighboring 
metal absorption transition.
%%%%
\begin{figure}
\psfig{figure=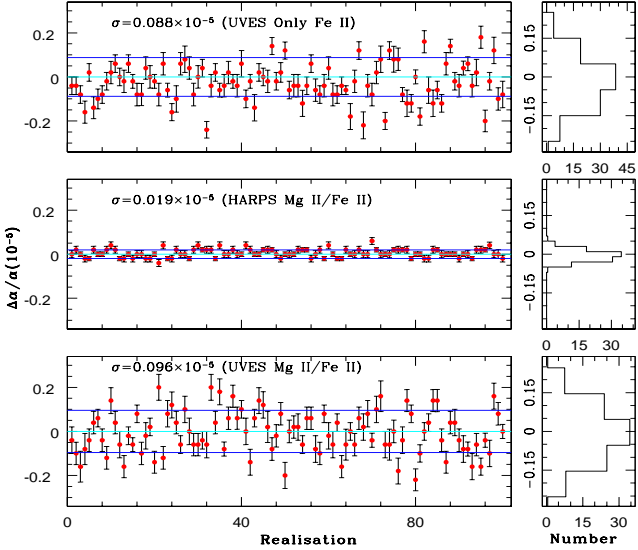,height=8.cm,width=9.cm,angle=00}
\caption[]{
Inferred \dela due to uncertainties in the wavelength calibration.
The results presented in the middle and bottom panels are when we
consider 5 main Fe~{\sc ii} lines along with the Mg~{\sc ii} doublet in our analysis.
For each realization a set of 21 Th-Ar lines are randomly chosen
(3 in the neighborhood of each 5 \feii and 2 \mgii lines of the \zabs~=1.1508 system).
\dela is computed from the measured deviations in the line
centroid by assigning the $q$ coefficient (Dzuba et al. 2002) of
neighboring metal line to the Th-Ar emission line. 
The left-hand side middle and lower panels
show the result for the  HARPS and UVES spectra respectively. 
The histogram for both cases are shown in the right-hand side
panels. The top panels gives the results for the UVES data
when we consider 6 Fe~{\sc ii} lines (i.e Fe~{\sc ii}$\lambda1608$ and 5 main
Fe~{\sc ii} lines) in the analysis.
The $\sigma$ of the distribution refers to
a typical error on the measurement of \dela due to wavelength
calibration alone in a single system with 3 distinct components.
}
\label{dalpha.fig} 
\end{figure} 
%%%

All the lamp emission lines in each realization are fitted  
simultaneously with Gaussians,
for one fixed value of \dela. Here, the \dela value is used to 
modify the rest wavelength of the emission lines using the $q$ 
coefficients given by  Dzuba et al. (2002) for the corresponding metal lines.
This procedure is repeated for a range of \dela, from
$-2.0\times10^{-5}$ to $2.0\times 10^{-5}$ in steps of $0.02\times10^{-5}$
to achieve \chisq as a function of \dela. The \chisq versus \dela curve is used 
to extract the best fitted \dela (with error-bars) in a similar way
as is used in the absorption system (discussed in the next Section). The
measured spurious \dela for 100 
random realizations are plotted in Fig.~\ref{dalpha.fig} both for
HARPS (left-hand side middle panel) and UVES (left-hand side lower panel) lamp spectra.
In the top panel we give the results for similar analysis of UVES spectrum 
considering 6 Fe~{\sc ii} lines (i.e including Fe~{\sc ii}$\lambda$1608 instead
of Mg~{\sc ii} doublet) alone. 
\par
We notice that the measured values of \dela obtained in this experiment 
have a Gaussian-shape distribution with $\sigma$~of~$0.02\times 10^{-5}$ for HARPS 
and $\sigma \simeq 0.1\times10^{-5}$ for UVES.
As the system under consideration
is known to have much more than 3 components, the above quoted values
are conservative errors 
due to uncertainties in the wavelength calibration. 
Murphy et al. (2003) have also carried out such analysis 
for HIRES Th-Ar lamp spectra. Their weighted mean from the 
sample of 128 sets of Th-Ar lines resulted in
$\langle\Delta\alpha/\alpha\rangle_{ThAr}=(0.4\pm0.8)\times10^{-7}$. If one 
assumes a Gaussian distribution for the individual values, then
the central limits theorem implies that the typical $\sigma$ from 
one set of Th-Ar lines in the case of HIRES should be around
$0.09\times10^{-5}$ ($\equiv0.8\times10^{-7}\times\sqrt{128}$), which is
similar to our value for UVES Th-Ar lamp spectra 
(i.e $\sigma = 0.1\times10^{-5}$).\par
%%%
\subsection{Effect of using different Th-Ar line tables on wavelength calibration}
Th-Ar reference wavelengths are taken from the compilations of  
Palmer et al. (1983) for Thorium lines and
Norl\'en et al. (1973) for Argon lines. The line lists
built from these compilations and commonly used for echelle
spectroscopy calibration are available on the web-pages of the European Southern
Observatory (ESO\footnote{http://www.eso.org/instruments/uves/tools/tharatlas.html})
and  the National Optical Astronomy Observatory
(NOAO\footnote{http://www.noao.edu/kpno/specatlas/thar/thar.html}).
The two tables differ slightly, because the 
ESO Th-Ar line table is not accurate up to 4 decimal  
places as is the case with NOAO Th-Ar line table.
%%%
\begin{figure}
\psfig{figure=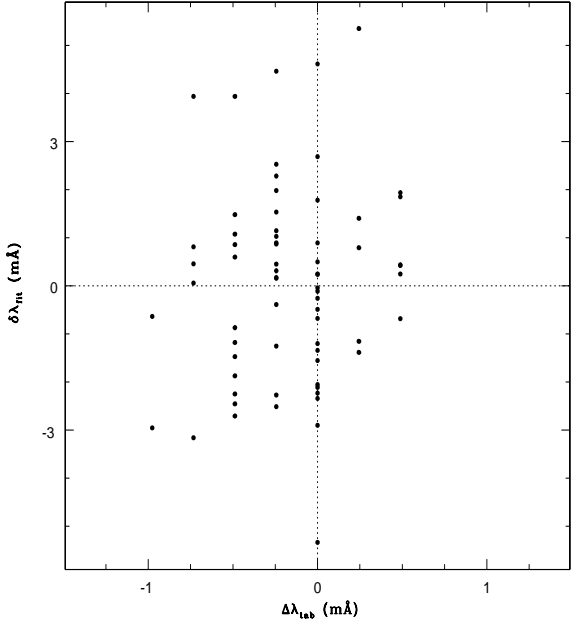,height=9.cm,width=8cm,angle=00} 
\caption[]{ The difference,  $\delta\lambda_{fit}$, between the best-fit 
centroid of Th-Ar lines seen in the UVES Th-Ar lamp spectrum
and their wavelength listed in the NOAO Th-Ar table is plotted versus
the difference between the corresponding wavelengths listed in the
ESO and NOAO Th-Ar line tables ($\Delta\lambda_{tab}$). 
The figure shows that (i) the scatter of
  $\Delta\lambda_{tab}$ is about a factor 3 smaller than that of $\delta\lambda_{fit}$ 
  (ii) no clear correlation is seen between $\Delta\lambda_{tab}$  and $\delta\lambda_{fit}$.
  As a result the calibration errors due to differences in wavelengths given
 in different Th-Ar tables 
is negligible as compare
  to the wavelength calibration accuracy of the instrument.}
\label{tab_diff.fig} 
\end{figure}
%%%%
For the extraction of UVES lamp spectra
we have used the Th-Ar line table provided by NOAO.
To investigate whether the use of ESO table 
could induce systematic shifts in \dela, we have also 
extracted the same UVES Th-Ar lamp
spectrum using the Th-Ar line table provided by ESO.
We fit a Gaussian function to the un-blended Th-Ar line as
described in sub-section~\ref{abscal:subsec} and get the deviation,
$\delta\lambda_{fit}$, of the best-fit centroid with respect to the 
corresponding value in the NOAO Th-Ar table. The deviation
($\delta\lambda_{fit}$) is plotted in
Fig.~\ref{tab_diff.fig} as a function of the difference in the
wavelengths tabulated by ESO and NOAO, $\Delta\lambda_{tab}$.
If the wavelength uncertainties caused by the inaccurate 
wavelengths listed in ESO Th-Ar table for some of the
Th-Ar lines are larger than the
errors allowed by the dispersion solution, then we expect a  
correlation between $\delta\lambda_{fit}$ and $\Delta\lambda_{tab}$.  
The lack of such a correlation and the larger scatter of 
$\delta\lambda_{fit}$ compared to $\Delta\lambda_{tab}$ in the 
figure, show that the effect of inaccurate rest-wavelengths of a few
lines in the ESO line list is negligible.\par
%%%%%
\begin{figure}[t]
\psfig{figure=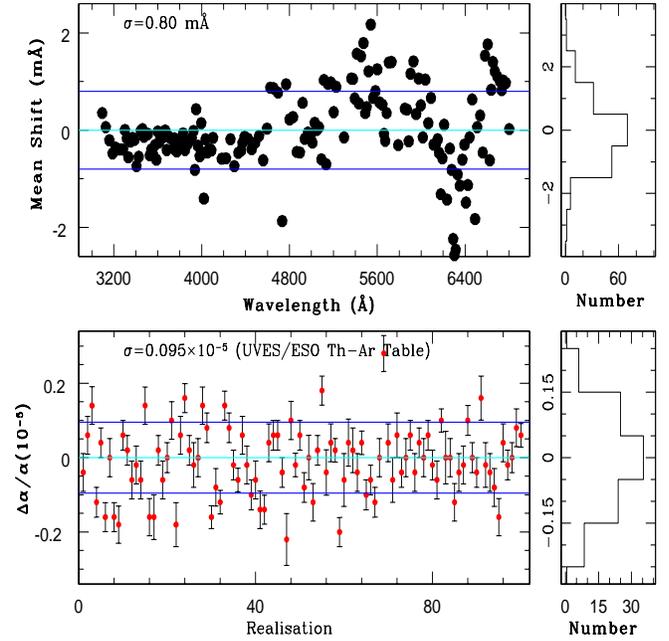,height=9.cm,width=9.cm,angle=00,bbllx=22bp,bblly=150bp,bburx=587bp,bbury=626bp,clip=true}
\caption{The figure shows the effect of calibration using  
 two different Th-Ar line tables:
  (i) provided by NOAO and usually used in IRAF (ii) provided by ESO and used in MIDAS.
  The upper left panel shows the mean shift of  Th-Ar lamp spectrum calibrated
 using ESO Th-Ar line table with respect to the same Th-Ar lamp spectra but 
 calibrated using the NOAO Th-Ar line table.
 The mean shift is 
 derived by performing  the cross-correlation as is shown in Fig.~\ref{crossc.fig} over a 
 wavelength region consisting of about 5 consecutive unblended Th-Ar lines.
 The histogram of the mean shift is shown in the right panel.
  The lower left panel shows the similar plot as in the 
  lower left panel of Fig.~\ref{dalpha.fig}, except that 
 here Th-Ar lamp spectrum is calibrated using the ESO Th-Ar lines table rather than
the NOAO Th-Ar lines table. The bottom right panel shows the histogram
of the \dela values.}
\label{crossNoaoEso.fig} 
\end{figure} 
%%%
To complement this,
we perform the cross-correlation between the lamp
spectra calibrated using the two wavelength tables. 
The cross-correlation is performed in a similar way as described
in sub-section~\ref{crossCorr:subsec}. 
Here we have shifted  the UVES lamp spectrum
calibrated using the ESO Th-Ar table over the same lamp spectrum  
calibrated using the NOAO Th-Ar line table. The result of the cross-correlation
is shown in the upper panel of Fig.~\ref{crossNoaoEso.fig}. From
the figure it can be seen that the relative shift is not completely random.
However the relative shift is most of the time less than 2m\AA~and even 1m\AA, which is 
well within the UVES calibration accuracy.\par
We also repeat the exercise to derive how these wavelength calibration
uncertainties translate into \dela  as described in detail in
sub-section~\ref{calDal:subsec} for the case when one uses 
for calibration the ESO Th-Ar line table (Fig.~\ref{dalpha.fig} for UVES lamp uses NOAO table). 
The result is shown in the lower
left-hand side panel of the Fig.~\ref{crossNoaoEso.fig} for 100 realizations. 
The histogram shown
in the lower right-hand side panel shows that the fiducial \dela is distributed
like a Gaussian. As a result, we can conclude that the \dela measurements in the
literature (Chand et al. 2004 \& 2005, Quast et al. 2004) using the ESO
Th-Ar line table, should not be significantly affected by this possible 
systematic effect.\par
%%%%%%%%%%%%%%%%%%%%%%%%%%%%%%%%%%SECTION 3 %%%%%%%%%%%%%%
\section{Analysis}
\label{sect:Ana}
In this section we present the results on the measurement of 
\dela using the HARPS and UVES spectra. The details of the analysis used
here, validation of the procedure using simulated spectra 
and the error budget from $\chi^2$ analysis can be 
found in Chand et al. (2004, 2005). Here, we mainly concentrate on 
(i) comparing the methods used by Chand et al. (2004, 2005) to derive \dela 
with that used by Quast et al. (2004) and (ii) understanding the effect of
the decomposition of the absorption profiles into multiple narrow Voigt-profile.
\par

%%%%
\subsection{Re-analysis of the red sub-system in the UVES data}
%%%%
%%%
\begin{figure*}
\psfig{figure=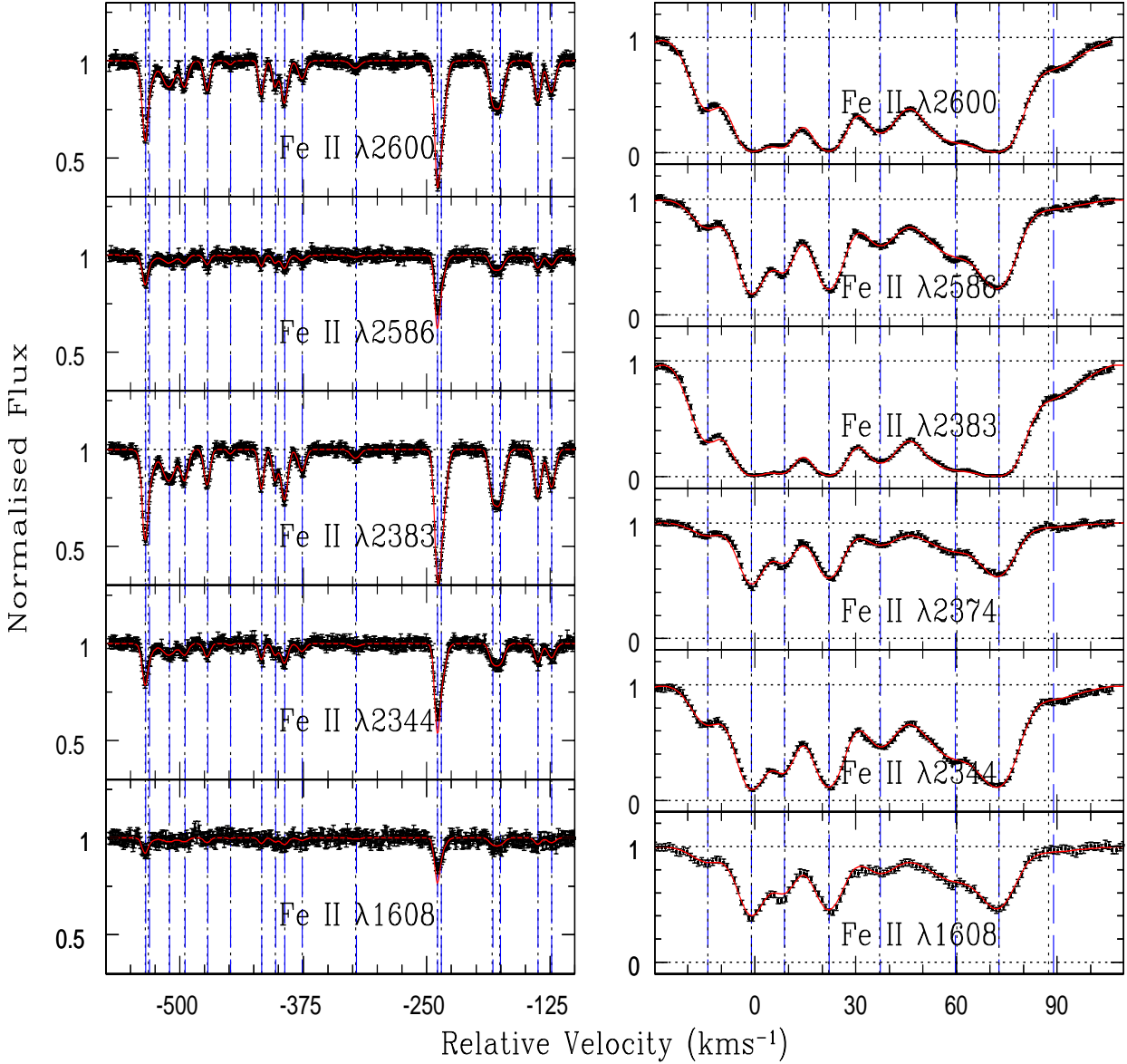,height=17.cm,width=17.8cm,angle=-90}
\caption[]{The figure shows the velocity plot of observed profiles
(data points with error-bars) together with the best fitted Voigt-profile
for \dela$=0$ over plotted as a solid curve, to the blue (left-hand side panels) and
red-subsystem (right-hand side panels) of the
\zabs~=1.1508 in the UVES spectrum. 
The dotted and dashed vertical lines are respectively the locations of the 
individual components obtained in this study and that of Quast et al. (2004).
}
\label{uves.fig} 
\end{figure*}
%%%%%%
\begin{figure*}
\psfig{figure=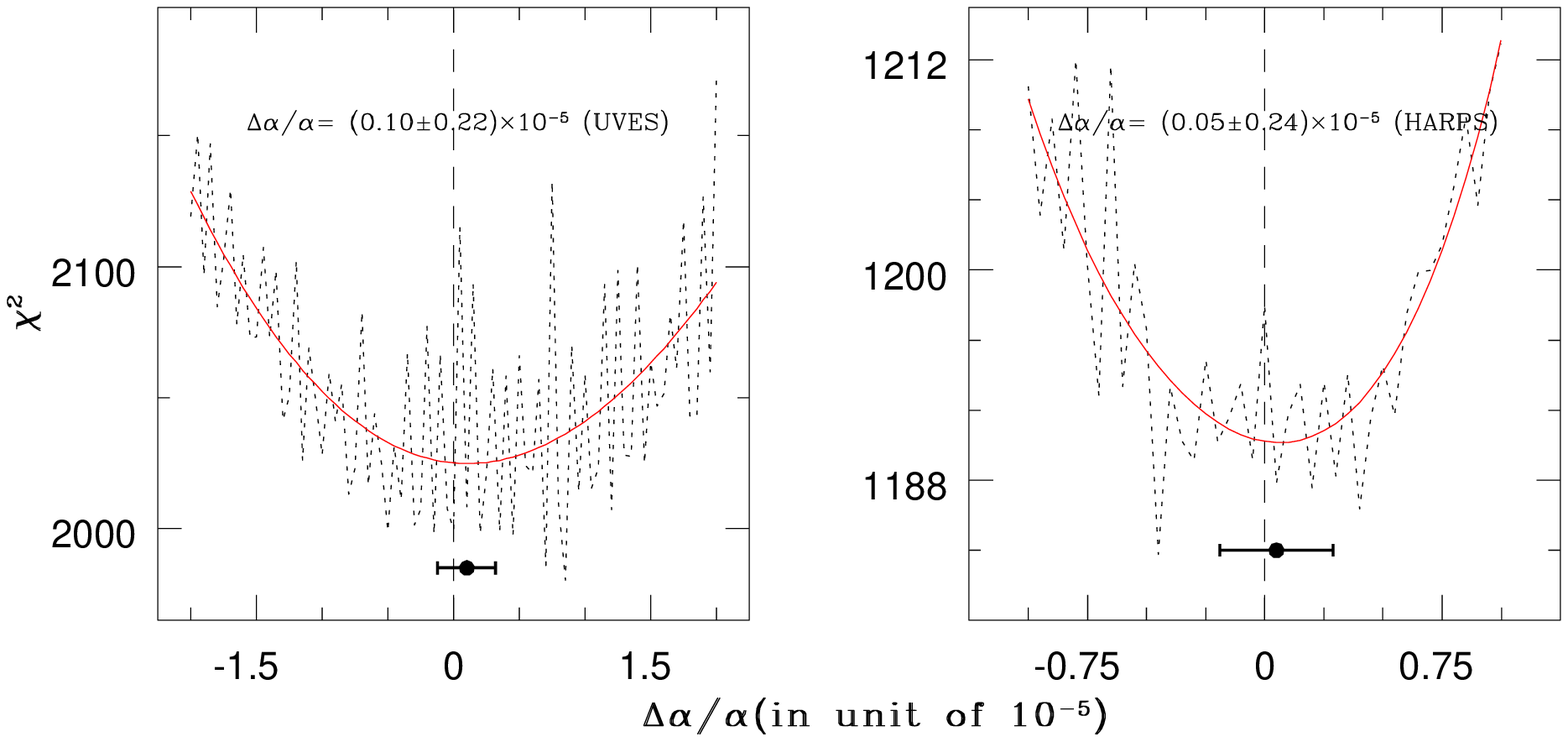,height=9.cm,width=17.8cm,angle=00,bbllx=22bp,bblly=430bp,bburx=579bp,bbury=698bp,clip=true} 
\caption{The dotted curve in both the panels indicates the variations 
of $\chi^2$ as a function of \dela as measured using the UVES (left panel)
 and HARPS (right panel) spectra.  The solid curve are the polynomial fit 
to these curve obtain using  rms minimization to avoid local fluctuations. Dark
rectangles with error bar indicate the position of the minimum
with one sigma error-bar obtained from $\chi^2_{min}+1$ statistics.
The \chisq curve in the left-hand side panel is derived based on the 
initial fit of UVES data shown in Fig.~\ref{uves.fig}, while 
the curve in the right-hand side panel is obtained by using
simultaneously the initial fit of HARPS data
shown in Fig.~\ref{fit_blueboth.fig} and 
Fig.~\ref{fit_redboth.fig}.}
\label{res.fig}
\end{figure*}
%%%%%
In the analysis of Chand et al (2004, 2005)
\dela is not explicitly used as fitting parameter. Instead \chisq versus \dela
curve is used to get the best fitted value of \dela. However, Quast et al.
(2004) use the Voigt profile analysis keeping \dela also as a fitting 
parameter in addition to $N$, $b$ and $z$. Chand et al. (2005), using
analytic calculations,  have shown that both the approaches should 
give the same result. Here we check this by re-analysing the  
absorption lines of the \zabs~=~1.1508 system toward HE {0515$-$4414}
using \chisq versus \dela curve.\par

The absorption lines of this system is spread over 
about 730 \kms (Quast et al. 2004). 
We have divided the whole system in two well  
detached blue and red sub-systems. The blue sub-system covers the velocity 
range $-$570 to $-$100 \kms  and 
the red sub-system covers the velocity range $-$20 to $+$110 \kms
with respect to \zabs=1.1508.
Our best fit Voigt-profiles to the blue and red 
sub-system using the UVES spectrum, is shown respectively in
the left and right-hand side panels of Fig.~\ref{uves.fig}.
The vertical dotted lines are best fitted velocity components
obtained in this study and the long dashed vertical lines mark 
the velocity components of the Quast et al. (2004). Apart form
the component around $\sim90$~\kms, we find almost perfect
matching between the components obtained
with two different fitting codes.
The variation of $\chi^2$ as a function
of \dela using  this initial fit (Fig.~\ref{uves.fig}) is shown
in the left-hand side panel of Fig.~\ref{res.fig}.
The scatter seen in the \chisq curve is mainly due 
to  low column density of many components in blue sub-system 
(see the discussion in Chand et al. 2004). 
The position of the minimum 
in the \chisq curve remains uncertain till either we smooth the curve
or fit some smoothing polynomial to it.  
Therefore we have fitted a polynomial function of 4th order minimizing the
rms deviation.
 The best fit of the \chisq curve is shown by the solid line
(left-hand side panel of Fig.~\ref{res.fig}).
 Its minimum gives 
\dela~=~${(0.10\pm0.22)\times10^{-5}}$,  using $\chi^2_{min}+1$ statistics. 
The derived  position of the minimum does not depart significantly
when we use a 2nd or 3rd order polynomial fit
to the $\chi^2$ data points.
Our best fitted value, \dela = ${(0.10\pm0.22)\times10^{-5}}$,
 is very much consistent
with that obtained by Quast et al. (2004)
(\dela~=~${[0.01\pm0.17]\times10^{-5}}$). 
\begin{figure*}
\psfig{figure=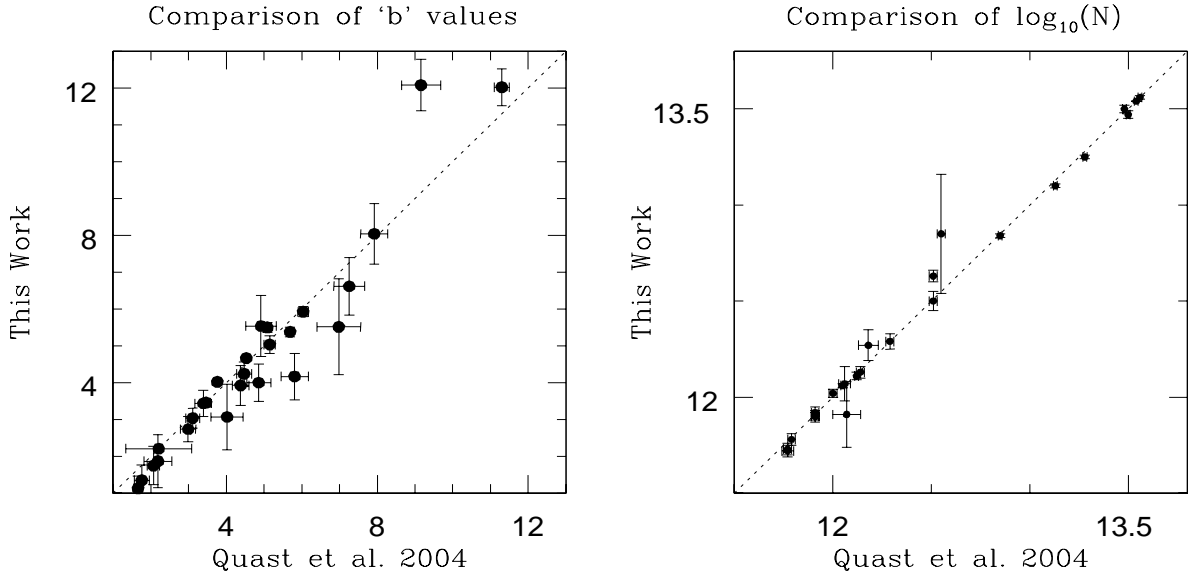,height=8cm,width=17.cm,angle=00,bbllx=27bp,bblly=430bp,bburx=600bp,bbury=716bp,clip=true} 
\caption{Comparison of parameters derived for individual components from the fit of the UVES spectrum
in this study and that of Quast et al. (2004).}
\label{compfit}
\end{figure*}
%%%
\begin{figure*}
\psfig{figure=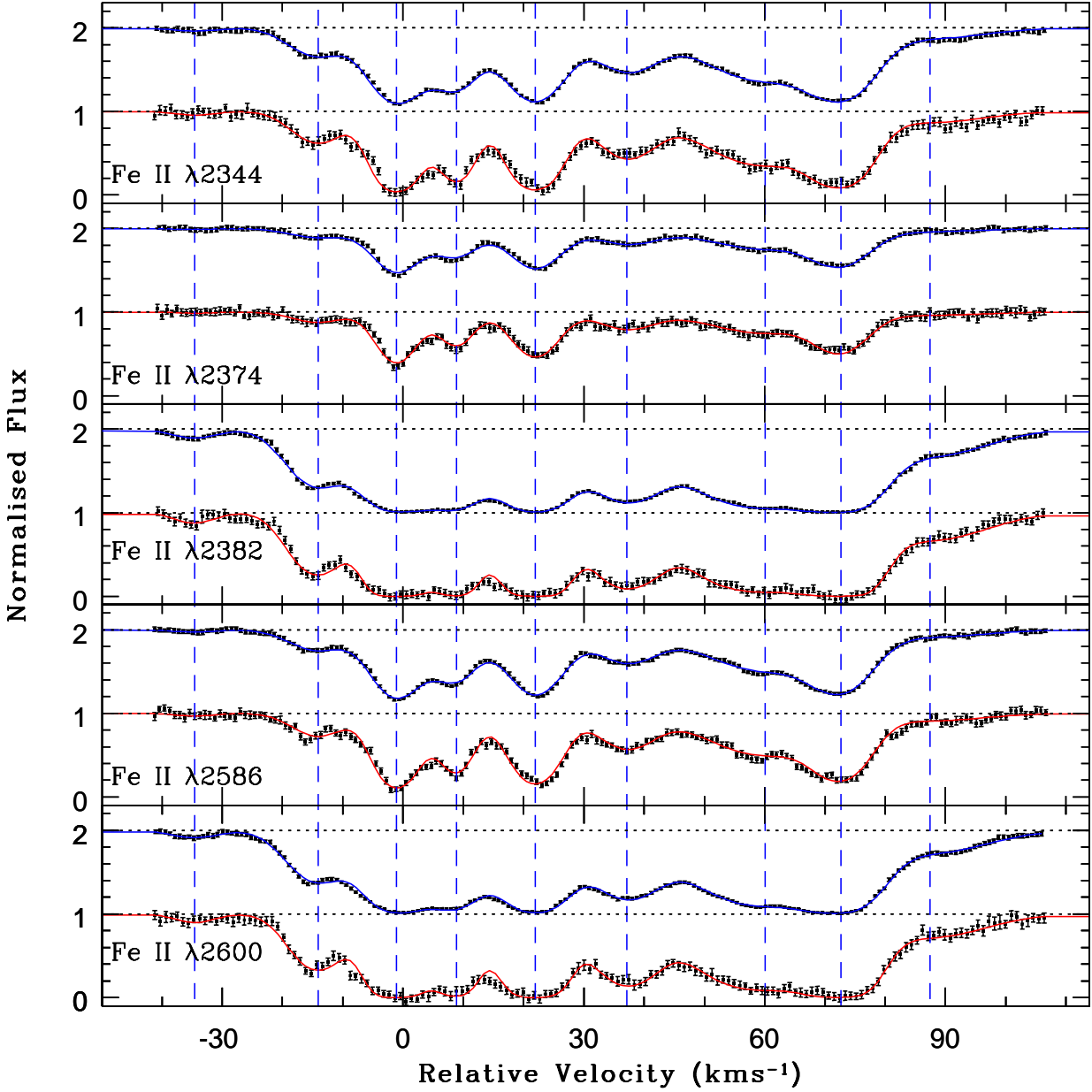,height=18.cm,width=17.7cm,angle=00}
\caption[]{Absorption profiles in the red sub-system of the
\zabs~=~1.1508 DLA toward HE~0515$-$4414 as observed with HARPS and
UVES plotted on a velocity scale. The normalized UVES spectrum is shifted 
in the y-direction by one unit for the sake of clarity.      
The data points with  error-bars correspond to the observed spectra. Over plotted 
as a solid curve is the best Voigt-profile fit based on the UVES data 
alone (same as in right-hand side panels of Fig.~\ref{uves.fig}).
For HARPS data the fit based on UVES data 
has been convolved
with HARPS instrumental profile. 
The figure demonstrates  the requirement for 
extra components, as evident from the higher spectral resolution HARPS 
spectrum (see for example region around $-$20 to 30 ~$kms^{-1}$).
}
\label{comp_stru.fig} 
\end{figure*} 
The best fitted column densities
and Doppler parameters in individual components also agree well 
(see Fig.~\ref{compfit}). 
%,
%%%
\begin{figure*}
\psfig{figure=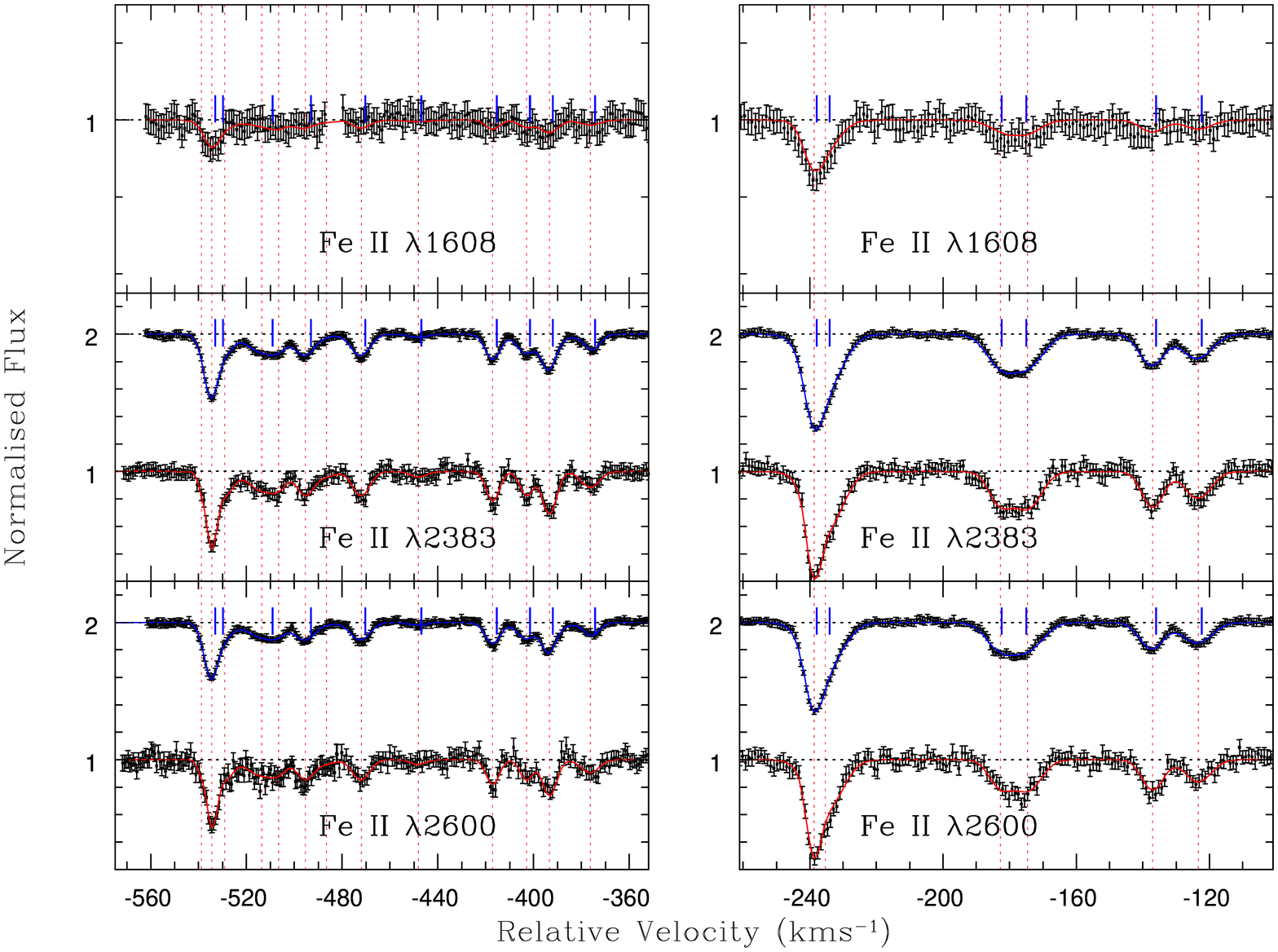,height=13.cm,width=18.7cm,angle=-90}
\caption[]{Absorption profiles in the blue sub-system of the \zabs~=1.1508
on a velocity scale.
The normalized UVES spectrum is shifted by unity for 
better visualization. The data points with  error-bars correspond to the 
observed spectra. 
Over plotted as solid curves are the best Voigt-profile
fits at \dela$=0$.
The Voigt-profile fits of both the HARPS and UVES data, shown here
are based on the component structure derived by imposing the condition that
the HARPS ($R=112000$) as well as the UVES ($R=55,000$ and better S/N) data
should be fitted with parameters that are consistent with each others.
The dotted vertical lines mark the
positions of components required
to fit the HARPS data. The thick ticks mark the position of
components as is derived using the UVES data alone (Fig.~\ref{uves.fig})
}
\label{fit_blueboth.fig} 
\end{figure*} 
%%%
%%%
\begin{figure*}
\psfig{figure=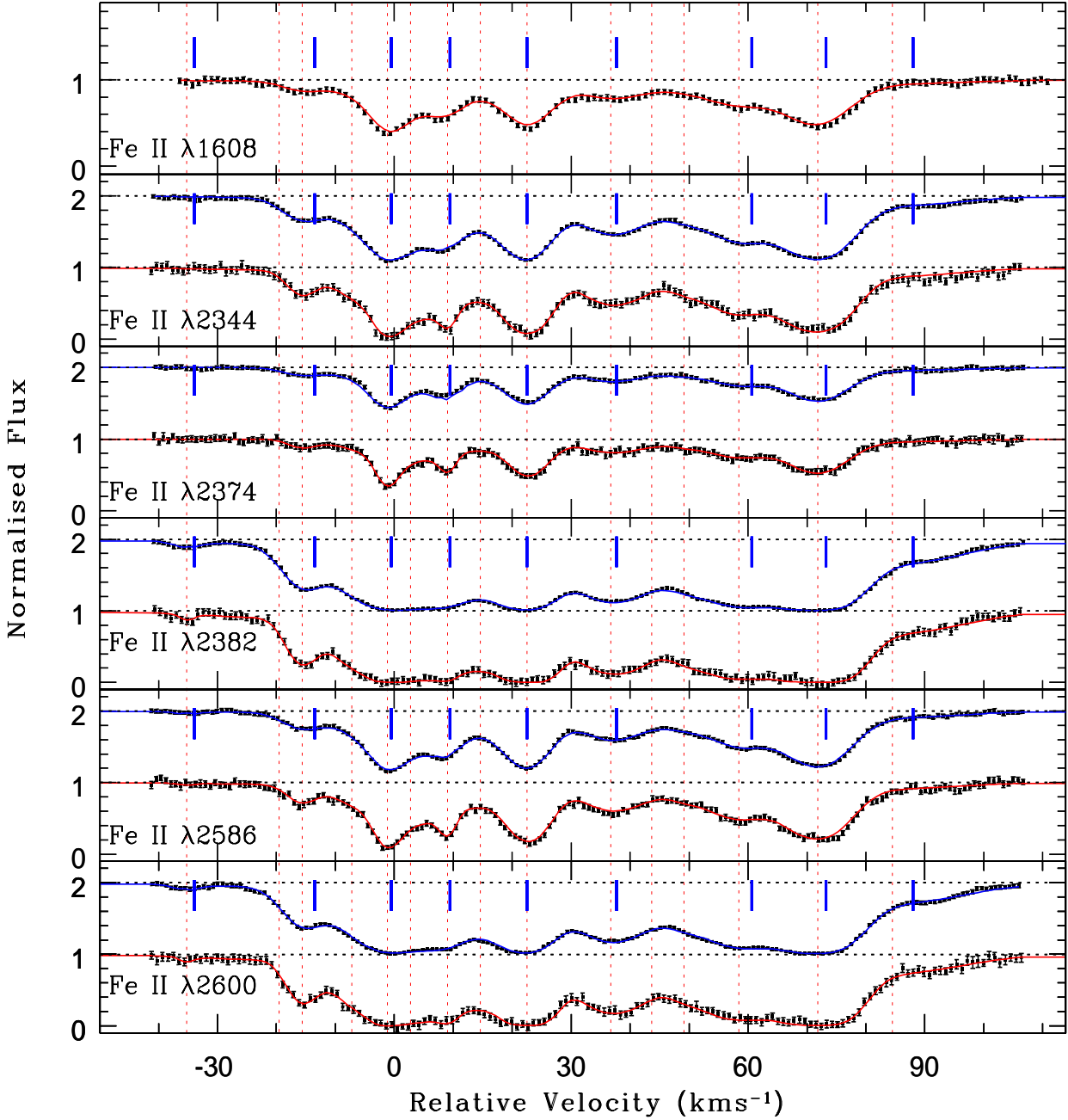,height=19.cm,width=17.8cm,angle=0} 
\caption[]{ Same as Fig.~\ref{fit_blueboth.fig} for the red sub-system. In addition, the figure
also illustrate more clearly that the HARPS data require
more components (15 components, shown by dotted line) compared to the 
UVES data alone (9 components, shown by thick ticks).
Also note that the fit shown for the UVES data is based on the 
component structure obtained in conjunction to HARPS data.
%These are not shown here for the sake of clarity,however 
For comparison the component required to fit the  UVES data alone 
(from Fig.~\ref{uves.fig} right-hand side panels) are marked by thick tick.}  
\label{fit_redboth.fig} 
\end{figure*}
    \begin{table*}
{\large
     \caption{Results of the Voigt profile fit of Fe~{\sc ii} lines at \zabs = 1.1508 toward HE {0515$-$4414}.}
     \begin{tabular}{rrrcr}
     \hline\hline
     \\
           {C.N}  & \multicolumn{1}{c}{\zabs}         &    \multicolumn{1}{c}{b} 
                  &     \multicolumn{1}{c}{log[N(Fe~{\sc ii})]}               &   \multicolumn{1}{c}{V$^{a}$} \\ 
                  &                                   & \multicolumn{1}{c}{(\kms)} 
	          & \multicolumn{1}{c}{(cm$^{-2}$)}& \multicolumn{1}{c}{(\kms)}    \\ 
     \\
     \hline\hline
     \\ 
  1 &$1.146938\pm0.00000^{\dagger}$  &$    1.70\pm0.22$	  &$   11.38\pm   0.14$  &$ -538.79\pm   00.00$ \\
  2 &$1.146969\pm0.000098$	      &$    2.34\pm0.25$          &$   12.30\pm   0.03$  &$ -534.46\pm   13.71$ \\
  3 &$1.147008\pm0.00000^{\dagger}$  &$    4.47\pm0.75$          &$   11.90\pm   0.06$  &$ -529.02\pm   00.00$ \\
  4 &$1.147117\pm0.001030$	      &$    7.45\pm1.01$          &$   12.01\pm   0.04$  &$ -513.80\pm   143.7$ \\
  5 &$1.147169\pm0.000410$	      &$    4.25\pm0.88$          &$   11.58\pm   0.09$  &$ -506.54\pm   57.27$ \\
  6 &$1.147249\pm0.000106$	      &$    4.63\pm0.22$          &$   11.92\pm   0.04$  &$ -495.37\pm   14.83$ \\
  7 &$1.147312\pm0.00000^{\dagger}$  &$    4.90\pm0.45$          &$   11.23\pm   0.17$  &$ -486.57\pm   00.00$ \\
  8 &$1.147416\pm0.000096$	      &$    4.70\pm0.19$          &$   11.93\pm   0.04$  &$ -472.05\pm   13.33$ \\
  9 &$1.147587\pm0.000255$	      &$    4.49\pm0.67$          &$   11.24\pm   0.15$  &$ -448.18\pm   35.65$ \\
 10 &$1.147809\pm0.000113$	      &$    3.47\pm0.22$          &$   11.91\pm   0.04$  &$ -417.19\pm   15.84$ \\
 11 &$1.147911\pm0.000133$	      &$    3.39\pm0.25$          &$   11.81\pm   0.04$  &$ -402.96\pm   18.58$ \\
 12 &$1.147980\pm0.000215$	      &$    3.75\pm0.57$          &$   12.12\pm   0.10$  &$ -393.33\pm   30.04$ \\
 13 &$1.148101\pm0.000543$	      &$    4.99\pm1.12$          &$   11.75\pm   0.07$  &$ -376.44\pm   75.84$ \\
 14 &$1.148501\pm0.000218$	      &$    7.52\pm0.44$          &$   11.56\pm   0.10$  &$ -320.62\pm   30.47$ \\
 15 &$1.148783\pm0.000287$	      &$    2.97\pm0.57$          &$   11.09\pm   0.18$  &$ -281.27\pm   40.03$ \\
 16 &$1.149088\pm0.000096$	      &$    2.11\pm0.21$          &$   12.44\pm   0.03$  &$ -238.72\pm   13.32$ \\
 17 &$1.149112\pm0.000057$	      &$    6.46\pm0.11$          &$   12.52\pm   0.03$  &$ -235.38\pm   07.97$ \\
 18 &$1.149489\pm0.000398$	      &$    4.30\pm0.43$          &$   12.03\pm   0.03$  &$ -182.79\pm   55.50$ \\
 19 &$1.149547\pm0.000470$	      &$    5.50\pm0.53$          &$   12.20\pm   0.02$  &$ -174.70\pm   65.55$ \\
 20 &$1.149817\pm0.000061$	      &$    4.14\pm0.12$          &$   12.08\pm   0.03$  &$ -137.05\pm   08.56$ \\
 21 &$1.149915\pm0.000108$	      &$    5.12\pm0.21$          &$   12.01\pm   0.03$  &$ -123.38\pm   15.02$ \\
 22 &$1.150548\pm0.00000^{\dagger}$  &$    0.26\pm0.0^{\ddagger}$&$   11.21\pm   0.16$  &$  -35.13\pm   00.00$ \\
 23 &$1.150659\pm0.00000^{\dagger}$  &$   17.85\pm0.0^{\ddagger}$&$   12.21\pm   0.06$  &$  -19.65\pm   00.00$ \\
 24 &$1.150688\pm0.000107$	      &$    2.98\pm0.18$          &$   12.58\pm   0.02$  &$  -15.61\pm   14.94$ \\
 25 &$1.150747\pm0.00000^{\dagger}$  &$    4.62\pm0.78$          &$   12.47\pm   0.32$  &$   -7.39\pm   00.00$ \\
 26 &$1.150792\pm0.000102$	      &$    1.95\pm0.18$          &$   13.26\pm   0.03$  &$   -1.11\pm   14.20$ \\
 27 &$1.150819\pm0.00000^{\dagger}$  &$    8.16\pm1.67$          &$   13.46\pm   0.07$  &$    2.65\pm   00.00$ \\
 28 &$1.150864\pm0.000126$	      &$    1.07\pm0.25$          &$   12.88\pm   0.05$  &$    8.92\pm   17.53$ \\
 29 &$1.150903\pm0.00000^{\dagger}$  &$    3.65\pm2.03$          &$   12.58\pm   0.37$  &$   14.36\pm   00.00$ \\
 30 &$1.150962\pm0.000190$	      &$    4.21\pm0.21$          &$   13.47\pm   0.03$  &$   22.58\pm   26.50$ \\
 31 &$1.151063\pm0.000207$	      &$    6.68\pm0.39$          &$   13.09\pm   0.02$  &$   36.66\pm   28.89$ \\
 32 &$1.151113\pm0.00000^{\dagger}$  &$    6.00\pm1.61$          &$   12.34\pm   0.11$  &$   43.63\pm   00.00$ \\
 33 &$1.151152\pm0.00000^{\dagger}$  &$    3.37\pm0.83$          &$   12.25\pm   0.08$  &$   49.06\pm   00.00$ \\
 34 &$1.151218\pm0.000235$	      &$    7.13\pm0.38$          &$   13.29\pm   0.02$  &$   58.26\pm   32.68$ \\
 35 &$1.151314\pm0.000158$	      &$    6.21\pm0.17$          &$   13.56\pm   0.02$  &$   71.64\pm   22.06$ \\
 36 &$1.151406\pm0.00000^{\dagger}$  &$   15.40\pm0.0^{\ddagger}$&$   12.72\pm   0.02$  &$   84.46\pm   00.00$ \\
 \\                                                                                        
 \hline 

\multicolumn{5}{l}{`$^{a}$' relative velocity with respect to $z_{\rm abs}=1.1508$.}\\ 
\multicolumn{5}{l}{`$^{\dagger}$' The redshift ($z$) of these components are kept fixed.}\\ 
\multicolumn{5}{l}{`$^{\ddagger}$' The Doppler parameter, $b$, of these components are kept fixed.}\\ 
\label{model.tab} 
\end{tabular}
}
\end{table*}                                                                           
 %%%%
The larger errors in the measured
quantities in the present study is mainly due to higher values of 
the error assigned to the flux in individual pixels.
 Thus the analysis presented here clearly shows that
the analysis used by us in Chand et al (2004, 2005) produces 
consistent results.\par
In addition we have also performed the analysis
of UVES spectra by excluding the weaker \feii lines
from the blue sub-system and heavily saturated 
strong \feiistr lines from the red-subsystem,
 (see discussion in Chand et al. 2004). 
In this case the \chisq curve is found relatively less 
fluctuating as compare to the left-hand side panel of Fig.~\ref{res.fig}, 
and has resulted in \dela = $(0.00\pm0.26)\times 10^{-5}$.
%%%
\subsection{\dela from the HARPS data}
The decomposition of the absorption profiles in sub-components is
expected to be better defined from the HARPS spectrum because 
of its superior spectral resolution.
In Fig.~\ref{comp_stru.fig} we compare the profiles of the
\feii lines in the red sub-system 
as observed with HARPS and UVES. The best multi-component 
Voigt-profiles fit using the UVES spectrum alone is over plotted.
To fit the HARPS data we need additional components, 
as is apparent in the region around $-$20 to $+$30 \kms where consistent
differences are seen for all profiles between the HARPS spectrum and the fit to the
UVES data alone. 
However, the UVES spectrum has the advantage of having higher S/N. Thus, 
in our analysis we fitted simultaneously
both HARPS and UVES data using the same component structure
and the appropriate instrumental functions.
We initially fitted the HARPS data and used the derived parameters to fit the UVES data. 
The process was repeated until the residuals along the profiles are symmetrically 
distributed around zero and the best-fit parameters from these two data 
sets are consistent with one another within measurement uncertainties. 
In this exercise we have not included the line \feiia (covered only in the UVES spectrum) 
so that our derived component structure is not artificially bias towards \dela$=0$.
\par
Our best-fit Voigt-profile components that simultaneously fit
the HARPS and UVES spectra are shown in Fig.~\ref{fit_blueboth.fig},\ref{fit_redboth.fig} 
respectively for the blue and red sub-systems. 
The best-fit parameters are listed in Table.~\ref{model.tab}. 
The component identification number (C.N), redshift ($z$), velocity
dispersion ($b$), and Fe~{\sc ii} column density ($N$), for each component
are listed respectively in columns 1, 2, 3 and 4. The last column of the table lists the 
relative velocity of the components with respect to \zabs=1.1508.
We find that the blue and red sub-system (Fig.~\ref{fit_blueboth.fig},\ref{fit_redboth.fig})
require respectively 3 and 6 extra components
compared to the minimum number required to
fit the UVES spectrum alone with $\chi^2=1$.\par
%%%
We evaluate the best-fit \dela value using the  high resolution HARPS 
spectrum for the five main Fe~{\sc ii} lines and the UVES spectrum for
Fe~{\sc ii}$\lambda1608$ considering  both the blue and red 
sub-systems simultaneously.
Here it should be noted that the
\feiia is crucial for \dela measurement due to its 
opposite sensitivity for \dela (negative $q$ coefficient) 
compared to the other main \feii lines. However as its 
observed wavelength range ($\approx 3460$\AA) is not 
covered by the HARPS spectral coverage (3800 - 6900\AA),  
we have to use it from the UVES spectrum for constraining the 
\dela value.
The $\chi^2$ versus \dela curve is shown in the 
right-hand side panel of Fig.~\ref{res.fig}.
The scatter seen in the \chisq curve is mainly due 
to the low S/N ratio and low column density of many components as can be seen from 
Table~\ref{model.tab} (see the discussion in Chand et al. 2004). 
The continuous curve gives the 4th order 
polynomial fit to the $\chi^2$ data points
using rms minimisation.
Its minimum gives 
\dela~=~${(0.05\pm0.24)\times10^{-5}}$,  using $\chi^2_{min}+1$ statistics. 
This result is consistent with the Quast et al. (2004) measurement 
(\dela~=~${[0.01\pm0.17]\times10^{-5}}$) based on the UVES spectrum and 
lesser number of components. Thus in this particular case lack of
information on the additional components in the UVES spectrum does not
seem to affect the final result. 
%%%%%%%%%%%%%%%%%%%%%%%%%%%%%%%%%%%%%%%%%%%%%%%%%%%%%%%%%%%%%%%
\section{Result and discussion}
\label{sect:result}
In this paper, we present a very high resolution (R~=~112,000) spectrum of
QSO HE 0515$-$4414 obtained using HARPS.
We have used the high wavelength calibration accuracy and high 
spectral resolution capabilities of HARPS to address the following
issues.\par
We compare the lamp spectra obtained with UVES and HARPS. 
Using cross-correlation analysis we show that any possible relative
shift between the two spectra are within 2~m\AA. 
Using Gaussian fits to unblended lamp emission lines, we find that the 
absolute wavelength calibration of HARPS is very robust with rms 
deviation of 0.87~m\AA~with respect to the wavelengths tabulated in Cuyper et al. (1998).
This is about a factor of 4 better than that of
UVES ($\sigma=4.08$ m\AA,~see Fig.~\ref{dlam.fig}).
Thus the small shifts noted between the HARPS and UVES lamp spectra
are well within the typical wavelength calibration accuracy of UVES.
We have derived the error on \dela measurements  
due to the calibration accuracy alone. For UVES and HARPS spectra this 
is found to be  respectively 
$\sigma=0.96\times10^{-6}$ and  $\sigma=0.19\times10^{-6}$ for a typical 
system with three well detached components.
The value obtained for the UVES spectrum is 
also consistent with that of HIRES (Murphy et al. 2003). 
\par
%%%%
This shows that HARPS is the ideal instrument for this kind of measurement.  
Unfortunately it is mounted on the 3.6~m telescope at La~Silla and 
only HE~0515$-$4414 is bright enough to be observed in a reasonable
amount of time. This shows as well that the UVES spectra reduced (or calibrated)
with the UVES pipeline and used in the literature to constrain 
 \dela (Srianand et al. 2004 and Chand et
 al. 2004, Quast et al. 2004, Chand et al. 2005)
 do not suffer from major systematic error in the
 wavelength calibration. \par
%%%%
We have obtained the accurate 
multi-component structure using the higher resolution data
(R~$\approx 112,000$ for HARPS compared to $\approx 55,000$
for UVES). The best fit to the profiles obtained by fitting
simultaneously the HARPS data (of higher resolution) and
the UVES data (of better S/N ratio) require additional 
components as compared to the fit using the UVES data alone
(Quast et al. 2004). Using this new sub-component decomposition
and both HARPS and UVES data, we find 
\dela~$=(0.05\pm0.24)\times10^{-5}$. This is 
consistent with the results derived by Quast et al. (2004) 
from the UVES data alone. Indeed, we have in addition
re-analyzed the UVES data which was used in 
Quast et al. (2004) (without using the component structure from HARPS data),
to estimate the effect of different 
independent algorithms used to obtain error spectra, to combine the data,
to fit the continuum and to fit the absorption lines. We find that the best-fit
parameters as well as the
\dela measurement (\dela~=~${[0.10\pm0.22]\times10^{-5}}$), 
obtained by our independent analysis, are consistent with  that of 
Quast et al. (2004) (\dela~=~${[0.01\pm0.17]\times10^{-5}}$).\par
%%%
We note that the precision on the \dela measurement obtained 
using the HARPS spectrum, which is of high resolution and low S/N ratio, 
is similar to that obtained from the UVES spectrum, which is of lower resolution 
and higher S/N ratio.
Therefore, the improvement in the wavelength calibration 
accuracy by an order of magnitude using  HARPS will be effective to 
improve the constrain on \dela  only if high S/N ratio can also be obtained. 
This could be 
possible if an instrument such as HARPS can be mounted on bigger
telescopes.
\section*{Acknowledgments} 
HC thanks CSIR, INDIA for the grant award 
No. 9/545(18)/2KI/EMR-I. RS and PPJ gratefully acknowledge support from the Indo-French 
Centre for the Promotion of Advanced Research (Centre Franco-Indien pour 
la Promotion de la Recherche Avanc\'ee) under contract No. 3004-3. 
PPJ also thanks IUCAA (Pune, India) for hospitality during the time part of this
work was completed. RQ has been supported by the DFG under Re353/48.
% 
%%%
%
 
%
% 

\begin{thebibliography}{}
%
\bibitem{} Bahcall, J. N., Sargent, W. L. W. \& Schmidt, M. 1967,  
ApJ, 149, L11 
% 
\bibitem{} Bahcall, J. N., Steinhardt, C. L., \& Schlegel, D. 2004,
  ApJ, 600, 520  
% 
\bibitem{} Chand, H., Srianand, R., Petitjean, P., Aracil, B., 2004,
  A\&A, 417, 853
\bibitem{} Chand, H., Petitjean, P, Srianand, R., Aracil, B., 2005, A\&A, 430, 47
\bibitem{} Chengalur, J. N., Kanekar, N., 2003, Phys. Rev. Lett., 91, 241302
% 
\bibitem{} Cowie, L. L., \& Songaila, A., 1995, ApJ, 453, 596 
%
\bibitem{} Cuyper, De- J.-P., Hensberge, H., 1998, A\&AS, 128, 409
%
\bibitem{} Darling, J., 2003, Phys. Rev. Lett., 91, 011301 
%
\bibitem{} Darling, J., 2004, ApJ, 612, 58 
% 
\bibitem{} De~la Varga, A., Reimers, D., Tytler, D., et al. 2000, A\&A,
  363, 69
%
\bibitem{} Dzuba, V. A., Flambaum, V. V., Kozlov, M. G. et al. 2002, Phys. Rev A., 66, 022501
%
\bibitem{} Edl\'en, B. 1966, Metrologica, 2, 71
%
\bibitem{} Fischer,~M et al., 2004, Phys. Rev. Lett., 92, 230802
%
\bibitem{} Fujii,~Y, et al., 2000, Nucl. Phys. B, 573, 377
% 
\bibitem{} Griesmann U., \& Kling, R., 2000, ApJ, 536, L113 
% 
\bibitem{} Kanekar, N., Chengalur, J. N.,  2004, MNRAS, 350, L17
% 
\bibitem{} Levshakov, S. A. 1994, MNRAS, 269, 339 
%
\bibitem{} Levshakov, S. A., Centuri\'on, M., Molaro, P., D'Odorico, S.	
 	A\&A  2005a, 434, 827
%
\bibitem{} Levshakov, S. A., Centuri\'on, M., Molaro, P., D'Odorico, S.,  Reimers, D., Quast, R., 
           Pollmann, M., 2005b, ArXive Astrophysics e-prints, astro-ph/0511765
%
\bibitem{} Marion,~H., Phys. Rev. Lett., 2003, 90, 150801
% 
\bibitem{} Martinez, A. F., Vladilo, G., \& Bonifacio, P. 2003, MSAIS, 3, 252 
% 
\bibitem{} Martin, W. C, Zalubas, R., 1983, J. Phys. Chem. Ref. Data, 12, 323 
% 
\bibitem{} Mosser, B., Michel, E., Samadi, R. et al. 2004, Messenger, 114, 20
%
\bibitem{} Murphy, M. T., Webb, J., Flambaum, V., Prochaska, J. X.,  \& Wolfe, A. M.  2001a, MNRAS, 327, 1237  
%
\bibitem{} Murphy, M. T., Webb, J., Flambaum, V., Drinkwater, M. J.,
  Combes, F., Wiklind, T. 2001b, MNRAS, 327, 1244
% 
\bibitem{}  Murphy, M. T., Webb, J. K., Flambaum, V. V. 2003, MNRAS, 345, 609 
%
\bibitem{} Norl\'en, G., 1973, Physica Scripta, 8, 249  
%
\bibitem{} Palmer, B. A. and Engleman, R., Jr., 1983, Atlas of the Thorium Spectrum, Los 
           Alamos National Laboratory
%
\bibitem{} Petitjean, P., \& Aracil, B. 2004, A\&A, 422, 523
% 
\bibitem{} Petitjean, P., Ivanchik, A., Srianand, R., et al., 2004,
  C. R. Physique, 5, 411
% 
\bibitem{} Potekhin, A. Y., \& Varshalovich, D. A. 1994, A\&AS, 104, 89 
% 
%
\bibitem{} Quast, R., Reimers, D., \& Levshakov, S. A. 2004, A\&A, 415, L7 
%
\bibitem{} Quast, R., Reimers, D., Smette, A. et al. 2005, Proceedings of the 22nd Texas Symposium on
  Relativistic Astrophysics at Stanford University, page 1416  
% December 13--17, 2004,
\bibitem{} Srianand, R., Chand, H., Petitjean, P., Aracil, B.,
  2004, Phys. Rev. Lett., 92, 121302  
%
\bibitem{} Tzanavaris, P., Webb, J. K., Murphy, M. T., Flambaum, V. V., Curran, S. J., 2005, Phys. Rev. Lett., 95, 041301
% 
\bibitem{} Uzan, J.-P, 2003, RvMP, 75, 403 
%
\bibitem{} Uzan, J.-P, 2004, ArXive Astrophysics e-prints, astro-ph/0409424
%
\bibitem{} Varshalovich, D. A., Panchuk, V. E. \& Ivanchik, A. V. 
1996, Astron. Lett., 22, 6
% 
\bibitem{} Varshalovich, D. A., Potkhin, A. Y. \& Ivanchik, A. V. 
2000, in Dunford R. W., Gemmel D.S., Kanter E. P., Kraessig B.,
Southworth S. H.,  
 Yong L., eds, AA Conf. Proc. 506, X-ray and Inner-shell Processes. Argonne 
 National Laboratory, Argonne, IL, 503 
% 
\bibitem{}  Webb, J. K., Murphy, M. T., Flambaum, V. V., Dzuba V.~A.,  Barrow
  J.~D.,  Churchill C.~W.,  Prochaska J.~X., Wolfe A.~M., 2001,
  Phys. Rev. Lett., {87}, 091301  
% 
\bibitem{} Wolfe, A. M., Brown, R. L., \& Roberts, M. S. 1976, Phys. Rev.  
Lett., 37, 177 
%
\end{thebibliography}
\end{document}